\documentclass[10pt,letterpaper]{article}
\usepackage[top=0.85in,left=2.75in,footskip=0.75in]{geometry}

\usepackage{amsmath,amssymb}

\usepackage{changepage}

\usepackage{textcomp,marvosym}

\usepackage{cite}

\usepackage{nameref,hyperref}

\usepackage[right]{lineno}

\usepackage[table]{xcolor}

\usepackage{array}

\newcolumntype{+}{!{\vrule width 2pt}}

\usepackage{siunitx}
\usepackage{subcaption}
\usepackage{tabularx}
\usepackage{array}

\usepackage{graphicx}
\usepackage{amssymb}
\usepackage{amsmath}
\usepackage{rotating}
\usepackage{booktabs}
\usepackage{hyperref}
\usepackage{setspace}
\usepackage{dcolumn}
\usepackage{tabularx}
\usepackage{threeparttable}
\usepackage{floatrow}
\usepackage{threeparttable}
\usepackage{longtable}

\usepackage{color,soul}

\usepackage{algorithm}
\usepackage{algpseudocode}

\newlength\savedwidth

\raggedright
\usepackage[aboveskip=1pt,labelfont=bf,labelsep=period,justification=raggedright,singlelinecheck=off]{caption}

\makeatletter
\renewcommand{\@biblabel}[1]{\quad#1.}
\makeatother

\usepackage{lastpage,fancyhdr,graphicx}
\usepackage{epstopdf}
\fancyheadoffset[L]{2.25in}
\fancyfootoffset[L]{2.25in}
\begin{document}
\vspace*{0.2in}

\begin{flushleft}
{\Large
\textbf\newline{Key predictors for climate policy support and
political mobilization: The role of beliefs and preferences} 
}
\newline

 Montfort Simon\textsuperscript{1,2,}*

 \bigskip
 \textbf{1} University of Bern, Institute for Political Science, 3102 Bern, Switzerland
 \\
 \textbf{2} Oeschger Centre for Climate Change Research, Bern, Switzerland
\bigskip

\textcurrency Current Address: Institute for Political Science/Oeschger Centre for Climate Change Research/Fabrikstrasse 8/3012 Bern/Switzerland 

* simon.montfort@unibe.ch

\end{flushleft}
\section*{Abstract}
Public support and political mobilization are two crucial factors for the adoption of ambitious climate policies in line with the international greenhouse gas reduction targets of the Paris Agreement. Despite their compound importance, they are mainly studied separately. Using a random forest machine-learning model, this article investigates the relative predictive power of key established explanations for public support and mobilization for climate policies. Predictive models may shape future research priorities and contribute to theoretical advancement by showing which predictors are the most and least important. The analysis is based on a pre-election conjoint survey experiment on the Swiss CO$_2$ Act in 2021. Results indicate that beliefs (such as the perceived effectiveness of policies) and policy design preferences (such as for subsidies or tax-related policies) are the most important predictors while other established explanations, such as socio-demographics, issue salience (the relative importance of issues) or political variables (such as the party affiliation) have \emph{relatively} weak predictive power. Thus, beliefs are an essential factor to consider in addition to explanations that emphasize issue salience and preferences driven by voters' cost-benefit considerations.


\thispagestyle{empty}
\listoffigures
\listoftables

\section*{Introduction}
Climate change is one of the most pressing issues that humanity is currently facing. To mitigate its impacts, national governments have pledged to limit the impact of global greenhouse gas emissions (GHG) to a temperature increase of less than 2\textdegree \space compared to pre-industrial levels through the Paris Agreement. At the domestic level, the successful adoption of ambitious policies that impose stringent incentives for climate change mitigation crucially depends on public support and political mobilization\cite{bedsworth2013climate, lockwood2013political, anderson2017public}. Public support --  voter endorsement of climate policies that mitigate climate change -- shapes the introduction of the latter. In turn, political mobilization can amplify or dampen public support as it involves active participation in political processes, such as the signing of petitions, demonstrating, donating\cite{fisher2019all} or participating in elections\cite{holbrook2005mobilization}. Understanding the factors that influence both public support and political mobilization is thus essential when introducing ambitious climate policies. Therefore, this article asks: \emph{What are the most important predictors of public support and political mobilization for ambitious climate policies?}

This study fills two important gaps in the literature. First, despite the compound importance of adopting ambitious climate policy, public support and political mobilization\cite{grosser2010public, stokes2016electoral} are predominantly studied separately. This hampers our understanding of the factors that shape the successful adoption of ambitious climate policy. In terms of the outcomes of elections and popular votes, political mobilization concerning contested political issues like climate change mitigation policies can be decisive. In some instances, a lack of mobilization by supporting coalitions may even prevent the adoption of more ambitious climate policies. For instance, in the aftermath of the popular vote on the Swiss CO$_2$ law, news media speculated that the lack of political mobilization of supporting voters (in terms of casting their votes for the respective ballot) was partly responsible for the rejection of the law. Because of the compound importance of public support and political mobilization, more research is required that together investigates these two factors.

Second, prior research has mainly focused on \emph{explaining} rather than predicting\cite{shmueli2010explain, cranmer2017can, beiser2018assessing} public support\cite{drews2016explains, stokes2016electoral, huber2020public} and political mobilization, such as voter turnout\cite{mildenberger2019households}. Such explanatory research has aimed at testing if there is evidence for the theorized causes of outcomes. Although valuable for advancing our understanding of causal relationships, explanatory research often does not allow for holistic assessments of established explanations \emph{relative} to each other. Holistic assessment with predictive machine learning models can contribute to theoretical advancement \cite{shmueli2010explain} and the prioritization of research agendas by assessing competing explanations at the macro-level\cite{shmueli2010explain, cranmer2017can}. In the rapidly growing social sciences\cite{callaghan2020topography}, whose research disciplines are epistemologically and ontologically heterogeneous in their orientation and assumptions\cite{victor2015climate}, this is especially important. 

To address these two important gaps, this study presents evidence generated using a \emph{predictive} machine-learning model. Predictive modelling strives to maximize the correct anticipation of the outcome category for new and unseen data rather than testing for causal relationships based on theoretical expectations. Machine-learning models are appropriate for investigating the relative predictive importance of established explanations because of their significant flexibility in identifying patterns and interactions in complex data that involve many different predictors. One advantage in the context of predictive modelling is that they can take potential interaction effects into account without requiring these to be specified a priori \cite{kern2019tree}. In contrast, the models often used for theory-testing in explanatory research, such as ordinary least squares models, assume linear relationships between variables, making it necessary to specify potential interactions (see, e.g., \cite{wooldridge2010econometric}). From the established machine-learning models, I choose a random forest model\cite{breiman2001random} because there is useful literature that discusses the strengths and weaknesses of different variable importance scoring methods for random forest models\cite{breiman1984friedman, strobl2007bias, ziegler2014mining, nembrini2018revival, sandri2008bias, wright2017ranger, loh2021variable, biecek2021explanatory, gromping2007relative, gromping2009variable} and because it performed slightly better in predicting support and mobilization compared to XGBoost\cite{friedman2001greedy}. I employed survey experimental data collected before the rejection of the Swiss CO$_2$ Act in a popular vote in June 2021 triggered by a referendum. This empirical setting with experimental data helped reduce social desirability bias in the survey responses. The survey experiment involved presenting attributes closely connected to the actual vote, thus allowing for the realistic measurement of support for the respective policies. Survey experimental conjoint designs have been demonstrated to closely replicate the outcomes of Swiss referenda\cite{hainmueller2015validating}. The measurement of the empirical data used for this study thus has high external validity.

The results show that voters' beliefs and preferences constitute an important leverage point for increasing the political feasibility of ambitious climate policies that align with the goals of the Paris Agreement through enhancing public support and political mobilization. Beyond existing findings that beliefs are a statistically significant explanation for support for ambitious climate policy on their own right\cite{ingold2011network, kammermann2018beliefs, huber2020public, stadelmann2021public, thaller2023perceived}, the results here show that they have the highest predictive power \emph{relative} to all other factors in the model. Beliefs are important not only in the literature on public support\cite{mildenberger2019beliefs, huber2020public, stadelmann2021public, thaller2023perceived} but also in the literature on advocacy coalitions (e.g. \cite{jenkins1994evaluating, ingold2011network, ingold2014drivers, jenkins2014belief, jenkins2018advocacy}) as they constitute the ``glue" that makes coalitions stable\cite{jenkins2014belief}. In contrast, other established explanations, such as individual-level issue salience as a potential channel through which beliefs and preferences may influence support (which features prominently in electoral theories) \cite{repass1971issue, page1983effects, krosnick1990government, fournier2003issue, belanger2008issue, miller2017origins}, is of relatively weak predictive power compared to beliefs and preferences. A promising avenue for future research is examining the extent to which political campaigns may enable changes in beliefs -- for instance, through the strategic ordering of climate policies into sequences\cite{meckling2015winning, meckling2017policy, pahle2018sequencing, fesenfeld2022policy} that initially create positive beliefs and subsequently foster the introduction of increasingly ambitious policies over time.

The article proceeds as follows: The following section introduces established explanations for public support and political mobilization from the literature. The third section outlines the methods that were employed, including a description of the case, data, machine-learning models, model selection, and variable importance scores. The fourth section describes the results in terms of the predictive power of the variables. The fifth section concludes with a discussion.  

\section{Literature Review: Established Explanations for Support and Mobilization}

Established explanations for public support and political mobilization for climate mitigation policies can be categorized into material and immaterial factors. Table \ref{tab:lit_overview} lists each of these explanations in the first column, provides a definition in the second column, and illustrates key mechanisms that operate in the relationship between the explanatory factor and public support or political mobilization in the third. Material factors are more tangible and have direct implications for voters' monetary costs and benefits. They include policy design preferences\cite{kachi2015climate, tobler2012addressing, fesenfeld2022policy}, tax stringency, such as the size of the respective tax\cite{klenert2018making}, and socio-demographic variables\cite{egan2017climate, beiser2018assessing}. Research in this field has shown that voter support decreases as monetary costs for voters increase\cite{diekmann2003green} but that wealthier people are generally more willing to pay for climate change mitigation\cite{meyer2010affluent}. Relatedly, research on socio-demographic variables has highlighted that high-income, highly educated, and females more strongly support ambitious climate policy\cite{meyer2010affluent, hornsey2016meta, beiser2018assessing}. Immaterial factors include less tangible, psychological and perception-related aspects\cite{egan2017climate, van2018psychological, beiser2018assessing, huber2020public}, such as beliefs\cite{blumer2018two, mildenberger2019beliefs, stadelmann2021public} -- for instance, about the perceived effectiveness of policies\cite{huber2020public}, or issue salience\cite{repass1971issue, page1983effects, krosnick1988role, krosnick1990government, fournier2003issue, wlezien2005salience, belanger2008issue, miller2017origins}, meaning the importance of climate and environment relative to other issues. 

Relatively recent research has drawn attention to cross-sectoral dynamics in policy processes due to the COVID pandemic\cite{fisher2021climate, bergquist2020combining, botzen2021lessons, drews2022climate}. Positive perceptions of how the government addressed the pandemic are associated with stronger climate policy support, mediated through general trust in politicians\cite{drews2022climate}. However, prior research has also shown that adverse impacts of health and economic conditions do not explain climate policy support. In the case of the COVID pandemic, this means that COVID concerns did not override climate issues and that the pandemic did not substitute support away from climate mitigation policies\cite{bergquist2020combining, drews2022climate}. One exception is unemployment due to COVID, which is negatively related to public support\cite{drews2022climate}.  

Another recently emerging strand related to the material and utilitarian-based dimensions of support and mobilization is termed policy sequencing, which is rooted in policy design, and the historical institutionalist theories of policy feedback, and path dependency\cite{montfortForthcoming}. Policy sequencing theory suggests that the strategic ordering of policies into sequences that first generate direct, policy-induced benefits for voters (for instance, through renewable energy subsidies) can later increase support for more cost-effective carbon pricing policies. Such carbon pricing policies are often advocated by economists to mitigate climate change as they are more economically efficient. However, the latter are often more weakly supported by voters because they generate direct and visible costs\cite{meckling2015winning, meckling2017policy, pahle2018sequencing, fesenfeld2022politics, montfortForthcoming}. Similarly, for identifying strategies that increase support for cost-effective carbon pricing policies, some policy design scholars argue that lump-sum reimbursement -- meaning that each voter receives an equal amount of revenue generated from a carbon pricing policy -- can increase support because this alters cost-benefit considerations\cite{klenert2018making}. However, others have emphasized that cognitive factors limit the impact of such benefits because the latter may remain unnoticed due to their limited visibility\cite{mildenberger2022limited}.

Seminal studies in economics have long highlighted the importance of material factors in shaping voter preferences based on utility-maximization\cite{downs1957economic, tullock1968mathematic, dhillon2002economic}. Downs\cite{downs1957economic} proposed that political turnout, a specific electoral component of political mobilization, is driven by the rational considerations of voters. Accordingly, voters weigh costs against benefits when deciding whether to vote or abstain. This implies that the benefits of voting are fundamentally driven by the anticipated probability of the impact of this activity. If voters do not expect their vote to have a decisive impact, they will abstain because the cost of voting is likely to outweigh the expected benefits, and vice versa. Overall, these considerations of individual voters determine voter turnout. According to this line of reasoning, beliefs are primarily associated with voters' rational appraisals of the utility of their activity\cite{dhillon2002economic}. 

Research on immaterial factors has shown that cognitive\cite{druckman2019evidence}, belief-\cite{stern1999value, ingold2011network, blumer2018two, kammermann2018beliefs, mildenberger2019beliefs, stadelmann2021public} and identity-based considerations\cite{shwom2010understanding, weber2010shapes, druckman2019evidence} explain a large proportion of the variation in policy support. Both voters and elites often evaluate policy proposals based on pre-existing, deep-held beliefs and convictions that simplify decisions\cite{kammermann2018beliefs}. For instance, theories about motivated reasoning suggest that voters (do not) update their opinions about specific issues because any new information runs counter to deeply held, identity-based beliefs\cite{druckman2019evidence}. This can amplify polarization along partisan lines\cite{unsworth2014s, shwom2010understanding}. Thus, identity-, belief-, and value-based opposition can make adopting policies more challenging. However, beliefs may also represent a channel for increasing public support for ambitious climate policies. The policy feedback literature suggests that the institutional and policy environment influences subsequent decisions\cite{soss2007public, beland2010reconsidering, campbell2012policy}. Positive public opinion feedback effects emerge for voters who believe climate change is real and trust political institutions\cite{stadelmann2021public}. 

\begin{table}[!htb]
\begin{adjustwidth}{-2.2in}{0in} 
\scalebox{0.9}{
\begin{tabular}{>{\hangindent=1em}p{1.25in}>{\hangindent=1em}p{2.2in}>{\hangindent=1em}p{4.25in}}
\toprule
 Concept & Definition & Key insights from established explanations \\
\midrule
\emph{Beliefs:} & Perceived effects of climate change mitigation policies & 
Beliefs that policies are effective at reducing CO$_2$ emissions and fair regarding the distributional effects increase public\cite{kronsik2006origins, huber2020public, gampfer2014individuals} and elite support\cite{ingold2011network, ingold2014drivers} because they create positive perceptions \\
\emph{Policy Design Preferences:} & Normative view of the types of policy that should be adopted & 
Policies that create financial support (e.g. subsidies) are more popular than policies that create direct costs (e.g. taxes) because the former generate direct revenues rather than direct costs\cite{ bemelmans2011carrots, tobler2012addressing} \\
\emph{COVID-Crisis:} & COVID-affectedness, government trust and satisfaction with crisis management & The COVID pandemic generally did not substitute support away from climate mitigation policies\cite{bergquist2020combining, drews2022climate} \\
\emph{Knowledge:} & Knowledge of voters about the content of climate policies & 
Knowledge is generally poor\cite{rhodes2014does, mildenberger2022limited}, but support increases when citizens are aware of policy content\cite{rhodes2014does} because there is less uncertainty about expected effects\\
\emph{Political Factors:} & Political left-right orientation and party affiliation & 
Decoupling partisan- and identity-based politicization from climate policy can reduce polarization because it decreases party cues \cite{van2018psychological, bliuc2015public }\\
\emph{Socio-demographics:} & Respondent attributes such as income, education, gender, or employment status & 
Socio-demographics have weak predictive power concerning climate attitudes\cite{beiser2018assessing}\\
\emph{Behavioural Factors:} & Aspects related to behaviour and low-carbon behavioural change & 
The increasing cost of behavioural change decreases climate policy support because voters weigh direct costs against benefits\cite{diekmann2003green}\\
\emph{Issue Salience:} & The importance of climate and environment relative to other issues, such as globalization & 
Issues of greater salience to voters shape their choices, especially when they interact with preferences because voters pay more attention to them compared to others \cite{repass1971issue, page1983effects, krosnick1988role, krosnick1990government, fournier2003issue, belanger2008issue, miller2017origins} \\
\emph{Tax Stringency:} & Specific stringency of carbon taxes (e.g. in the road and aviation transport, housing and food sectors) & Higher taxes decrease support\cite{carattini2019win, mildenberger2022limited, montfort2022sequencing} \\
\bottomrule
\end{tabular}}
\caption[Established explanations for public support and political mobilization]{Established explanations for public support and political mobilization.}
\label{tab:lit_overview}
\end{adjustwidth}
\end{table}

Among the immaterial explanations, issue salience (the importance awarded to climate and environment relative to other issues) is a prominent explanation for policy change. Salience may be conceptualized and operationalized on the individual or societal level\cite{miller2017origins}. The societal level includes public attention paid to an issue in media systems\cite{lax2012democratic}. On the societal level, normative theories of democracy stipulate that a shift in salience should be reflected in government policies and lead to policy change\cite{dahl1956preface, page1983effects}. The pluralist interplay in policy processes creates so-called ``polyarchic" systems\cite{dahl1956preface} in which governmental actors are responsive to the preferences of citizens\cite{schaffer2022policymakers}. Politicians may not be aware of public preferences regarding issues that lack salience for voters \cite{druckman2006lumpers, lax2012democratic} which may manifest in media attention to issues \cite{epstein2000measuring, lax2012democratic}. Similarly, other policy process theories (e.g. \cite{kingdon1984agendas}) posit that an increase in the salience of a problem can create a window of opportunity for leaders (so-called policy entrepreneurs) to invoke policy change because the public is more attentive to the issue\cite{zohlnhofer2016bringing}. 

At the individual level, research shows that voting behaviour is driven by the importance ascribed to issues\cite{krosnick1990government}. Voters evaluate politicians' performance more vigorously in relation to issues that they perceive as important\cite{fournier2003issue}. Political parties often mobilize based on issues that are salient to voters since they are more likely to evaluate the performance of the former based on these issues\cite{belanger2008issue}. Research on individual-level issue salience has used `most-important-problem' or `most-important-issue' questions in public opinion surveys \cite{wlezien2005salience, jennings2011distinguishing}. Explanatory research shows that greater issue salience often fosters policy change \cite{krosnick1990government, fournier2003issue, belanger2008issue}, including in the domain of climate policies\cite{bromley2020importance}. Although survey respondents may conflate most-important-issue and most-important-problem questions and thus there may be conceptual differences between the two question types\cite{wlezien2005salience}, results do not vary substantially between operationalizations. This is because respondents make similar cognitive associations when they are presented with (one of) the two question types\cite{jennings2011distinguishing}. In sum, evidence suggests that political support and mobilization, which shape policy change, are driven by issue salience. However, results may depend on whether aggregate-level data, such as media salience, or individual-level survey data, such as most-important-issue questions, are used \cite{miller2017origins} in interaction with other explanatory factors or in their own right. 

\section{Materials and Methods}

The following subsections describe the case of the Swiss CO$_2$ Act, the survey-experimental data, the machine-learning models, the model performance, and the variable importance scores used to analyze the relative importance of the predictors.  

\subsection{Case: Swiss Climate Mitigation Policy}

The Swiss CO$_2$ Act is the core instrument of Swiss climate policy. It was enacted in the year 2000 and revised in 2008 and 2013. In 2021, the Swiss population rejected a fourth revision that had previously passed through parliament. Switzerland is a direct democracy: decisions made by parliament are made subject to popular vote if 50,000 citizens sign a petition within 100 days calling for this. The proposed amendments to the act were rejected by a majority of 51.6\% on June 13, 2021.

To achieve net-zero GHG emissions by 2050, the 2021 revision of the CO$_2$ Act aimed to ratchet up the ambition of the reduction target and the stringency of the respective policy instruments. It included a more ambitious GHG reduction target of 50\% by 2030 compared to 1990 and carbon-pricing policy instruments in the housing, road transport, and aviation transport sectors. In the road transport sector, the CO$_2$ Act from 2013, which is still in place because the proposed act was rejected, mandates carbon pricing through an obligation for fuel importers to compensate for GHG emissions from imported fuels. This compensation was planned to be increased from 5 to 12 cents per litre in 2025\cite{parlament2020} corresponding to around 51 Swiss Francs per ton of CO$_2$. 

The act was supposed to extend the sectoral scope to the aviation transport sector, foreseeing a tax on private flights of at least 30 and maximally 120 Swiss Francs. In the housing sector, the act proposed a tax of 210 Swiss Francs per ton of CO$_2$ corresponding to an extra charge of roughly 55 cents per litre of heating oil. 

The revenues were intended to be used for two primary purposes. First, the CO$_2$ Act foresaw the creation of a climate fund to support firm-level investment into climate protection and renewable infrastructure, such as electric vehicle charging stations, in addition to supporting climate change adaptation in highly vulnerable alpine regions and villages. The fund should have been financed using one-third of the revenue from the carbon pricing policy instruments in the housing sector and maximally half of the revenue from the aviation transport sector. Such a creation of a fund to support the attainment of climate targets is also called revenue earmarking. Second, two-thirds of the revenue from the tax on fossil heating and at least half of the carbon tax in the aviation sector would have been reimbursed through lump-sum transfers to citizens, meaning that each citizen would have received an equal amount of money in the form of a credit on their old age insurance bill. 

\subsection{Data}

The data stems from a pre-election conjoint survey experiment. Such conjoint survey experiments have been demonstrated to closely replicate the outcomes of Swiss referenda\cite{hainmueller2015validating}. This experimental set-up thus has high external validity regarding actual voting behaviour. Moreover, since many of the key attributes in the CO$_2$ Act closely resemble the most important dimensions of the bill on which the public voted, this empirical setup allows for the precise measurement of public support. 

\begin{table}[!htb]
\scalebox{0.9}{
\begin{tabular}{>{\hangindent=1em}p{1.0in}>{\hangindent=1em}p{1.8in}>{\hangindent=1em}p{2.5in}}
\toprule
Concept & Question & Answer Categories \\
\midrule
\emph{Support:} & How strongly do you support or oppose each proposal? & 
\vspace{-.6cm}\begin{itemize}
	\setlength\itemsep{-.4em}
    \item Oppose fully
    \item Oppose
    \item Neither nor
    \item Support
    \item Support fully
    \vspace{-.2cm}
\end{itemize} \\
\emph{Mobilization:} & How much effort would you personally make to politically support or oppose the particular proposal (e.g., sign a petition, start a petition, demonstrate, donate, etc.)? & 
\vspace{-.6cm}\begin{itemize}
	\setlength\itemsep{-.4em}
    \item Much effort to prevent
    \item Some effort to prevent
    \item Neither nor
    \item Some effort to support
    \item Much effort to support
    \vspace{-.2cm}
\end{itemize} \\
\bottomrule
\end{tabular}}
\caption[Dependent variable question wording and answer categories]{Dependent variable question wording and answer categories. The survey data and question wording was translated from a conjoint survey experiment in Switzerland\cite{co2data}.}
\label{tab:dv_questions}
\end{table}

The survey was fielded precisely one week before the public vote on the Swiss CO$_2$ Act on June 13, 2021. It was terminated on June 17, 2021\cite{co2data}. The empirical study used questions about key variable groups discussed in the literature (see, e.g. Table \ref{tab:lit_overview}). The interactive tool that accompanies this paper to aid in the presentation of results (link: \url{https://prediction.adibilis.ch}) provides an overview of the wording of the survey questions and answers. These sum to 83 predictors in the model (see Table A, Fig A, and Fig B in S1 Text for further details). 

The survey was designed using the software Qualtrics in French and German and was fielded by the same provider. First, respondents were informed about the content of the survey and the anonymity of their responses. Second, study participants were randomly sampled with a quota restriction (also referred to as natural fallout) based on eligibility to vote, age, and gender to obtain a representative sample of the Swiss population. Voters below 18 years old or non-Swiss are ineligible to vote and were screened out, meaning they did not complete the entire survey. Respondents who wanted to complete the survey on their mobile phones were also screened out to ensure that the options from the conjoint survey experiment could be displayed clearly. Third, the survey contained general questions on salience and voting intentions. Fourth, these were followed by a block of questions on beliefs and preferences. Fifth came the survey-experimental conjoint questions aimed at measuring the effect of the tax stringency (the level of the tax) on support and mobilization, as described in more detail below. These were followed by questions on behavioural factors. Next came questions about COVID-19. Demographic questions were presented last to avoid raising the awareness of respondents about their socio-demographic identity. Median response duration was 19 minutes\cite{co2data}.

Table \ref{tab:dv_questions} shows the questions associated with the dependent variables in the fourth block of the survey experiment. The survey employed a bidirectional five-point Likert scale with a neutral middle category (this allowed respondents to express negative and positive stances). The authors of the survey opted for a five-point scale to limit the cognitive burden on the survey respondents compared to a seven-point alternative. The questions for the two dependent variables were presented after showing the respondents a randomized conjoint survey experiment. These six important features of the CO$_2$ Act captured in the conjoint survey experiment were debated extensively in parliament and among the public. Such features are called conjoint attributes and are summarized in Table \ref{tab:cjoint_attributes} alongside their randomized tax stringency levels, called conjoint attribute levels. Table \ref{tab:cjoint_attributes} relates these attributes from the conjoint experiment to the proposed act on which the population voted. Overall, the average policy proposal the respondents were asked to evaluate was more ambitious than the proposed act. 

In this experimental survey set-up, each respondent was shown a pair of policy proposals with randomized tax calibration attributes such as the reduction target, the size of the tax associated with housing, road transport, aviation transport, and the food sector, and the use of the revenue with a randomized attribute level from Table \ref{tab:cjoint_attributes}. Each respondent then rated the policy proposal by answering the questions in Table \ref{tab:dv_questions}. This procedure was repeated four times on a random sample of 1,511 Swiss voters. 

This random sample of 1,511 voters is representative of the population of Swiss voters: According to the method for sample size calculation by Yamane, the margin of error for the non-experimental questions is 0.0257, given the population of 5.5 million Swiss voters. This value is quite low, suggesting that the values observed in the sample are similar to those of the population (for details, see Text A in S1 Text). The total number of observations was $1,511$ (number of respondents) $\times 2$ (question regarding how much the respondent supports each proposal) $\times 4$ (how often conjoint survey-experiment was shown to each respondent) $= 12,088$. 

\begin{table}[!htb]
\begin{adjustwidth}{-2.2in}{0in} 
\begin{tabular}{>{\hangindent=1em}p{.8in}>{\hangindent=1em}p{4.2in}>{\hangindent=1em}p{1.9in}}
\toprule
Attribute & Randomized attribute levels & Level in the proposed CO$_2$ Act \\
\midrule
Reduction target & \vspace{-.5cm} \begin{itemize}
  \setlength\itemsep{-.2em}
    \item 40\% 
    \item 50\%  
    \item 60\% 
    \item 70\% 
    \item 80\%
\end{itemize} & 50\% GHG reduction target by 2030 relative to 1990\\
Carbon tax road transport & \vspace{-.45cm} \begin{itemize}
  \setlength\itemsep{-.2em}
    \item No tax on petrol (0 CHF/ton of CO$_2$)
    \item 0.14 CHF/l petrol (60 CHF/ton of CO$_2$)
    \item 0.28 CHF/l petrol (120 CHF/ton of CO$_2$)
    \item 0.42 CHF /l petrol (180 CHF/ton of CO$_2$)
    \item 0.56 CHF/l petrol (240 CHF/ton of CO$_2$)
\end{itemize} & 0.10 CHF/l petrol (43 CHF/ton of CO$_2$) in 2024 and 0.12 Fr./l petrol (51 CHF/ton of CO$_2$) in 2025 \\
Carbon tax housing  & \vspace{-.45cm} \begin{itemize}
  \setlength\itemsep{-.2em}
    \item No tax on heating oil (0 CHF/ton of CO$_2$)
    \item 0.16 CHF/l heating oil (60 CHF/ton of CO$_2$)
    \item 0.31 CHF/l heating oil (120 CHF/ton of CO$_2$)
    \item 0.47 CHF/l heating oil (180 CHF/ton of CO$_2$)
    \item 0.63 CHF/l heating oil (240 CHF/ton of CO$_2$)
\end{itemize} & 0.55 CHF/l heating oil (210 CHF/ton of CO$_2$) \\
Carbon tax food & \vspace{-.45cm} \begin{itemize}
  \setlength\itemsep{-.2em}
    \item No tax on meat (0 CHF/ton of CO$_2$)
    \item 0.77 CHF/kg meat (60 CHF/ton of CO$_2$)
    \item 1.53 CHF/kg meat (120 CHF/ton of CO$_2$)
    \item 2.30 CHF/kg meat (180 CHF/ton of CO$_2$)
    \item 3.07 CHF/kg meat (240 CHF/ton of CO$_2$)
\end{itemize} & The carbon tax on food was not included in the CO$_2$ Act on which the population voted\\
Carbon tax aviation transport & \vspace{-.45cm} \begin{itemize}
  \setlength\itemsep{-.2em}
    \item No tax on flights
    \item 10 CHF for short- and 30 CHF for long-distance
    \item 25 CHF for short- and 75 CHF for long-distance
    \item 40 CHF for short- and 120 CHF for long-distance
    \item 55 CHF for short- and 165 CHF for long-distance
\end{itemize} & Maximally 30 CHF for short- and 120 CHF for long-distance flights\\
Carbon tax revenue use & \vspace{-.45cm} \begin{itemize}
  \setlength\itemsep{-.2em}
    \item Exclusively lump sum reimbursement
    \item Mostly lump sum reimbursement
    \item Lump sum reimbursement and investment into climate protection
    \item Mostly investment into climate protection
    \item Exclusively investment into climate protection
\end{itemize} & Two-thirds lump sum reimbursement of the carbon tax heating oil, and half of the carbon tax in the aviation transport sector\\
\bottomrule
\end{tabular}
\caption[Conjoint attributes and attribute levels]{Conjoint attributes and randomized attribute levels and their relation to the proposed CO$_2$ act on which the public voted. The survey data and question wording were translated from a conjoint survey experiment in Switzerland\cite{co2data}.}
\label{tab:cjoint_attributes}
\end{adjustwidth} 
\end{table}

Assessments of public support and political mobilization depend on the policy proposals, meaning the value of the attributes, employed in conjoint survey experiments as well as other considerations by voters such as beliefs and preferences. While the former are conventionally used to estimate the causal effect of a set of policy attributes\cite{green1971conjoint, hainmueller2014causal, hainmueller2015validating}, the analysis described in this article additionally used non-experimental survey data to operationalize the other predictors that were derived from the literature (see Table \ref{tab:lit_overview}; for an overview of the survey questions and the respective answer categories, please consult the \url{https://prediction.adibilis.ch}). The use of an experimental outcome was associated with two advantages. First, because the conjoint survey experiment was repeated four times for two packages in each round, the number of observations was higher than when using non-experimental standard survey questions, meaning that statistical power was increased. This increased model fit. Second, the experimental setting reduced the social desirability bias that often plagues non-experimental outcomes\cite{horiuchi2022does}. Thus, the approach generated more reliable results than when applying a non-experimental outcome.

\subsection{Machine-Learning Model}

Machine-learning models support prediction because these computer algorithms are highly flexible in learning about relationships in the data that other models might miss. Many machine-learning models can, for instance, handle non-linear relationships in data, thus capturing potential interactions between predictors. For example, greater perceived fairness of climate policies regarding their distributive effects in society will increase support for ambitious climate policy. This relationship may be even stronger if climate change is an important issue for voters, the latter phenomenon is described as issue salience (see Table \ref{tab:lit_overview}). That is, two variables may have non-linear interaction effects, meaning that the magnitude of the effect of one variable depends on the value of another. Such interaction effects can naturally be captured with machine-learning models\cite{gromping2009variable}. Similarly, some machine-learning models, such as the random forest model used here, can also incorporate categorical predictors\cite{friedman2001greedy}. This is not the case for conventional statistical models often applied in explanatory research used for theory testing. For instance, one such model, the ordinary least squares model, assumes that the relationship between independent and dependent outcome variables is linear (see, e.g., \cite{wooldridge2010econometric}). Because of their flexibility, machine-learning models can use more information from the data to predict outcomes and hence provide better predictions which can be investigated using model performance statistics.

To compare model performance, I used two popular machine-learning models, the random forest\cite{breiman2001random} and the XGBoost model\cite{friedman2001greedy}. Fast implementations are available in \cite{wright2017ranger} for the random forest model and in \cite{chen2015xgboost} for the XGBoost model. These models permit treating the dependent variable, summarized in Table \ref{tab:dv_questions}, as a categorical outcome. Such models with categorical outcomes are also called supervised multi-class classification models. This permits more granular insight into the specific outcome categories than when treating the outcome as continuous. Each class represents a category of answers to the questions in Table \ref{tab:dv_questions}. This allowed for predicting the probability that a voter would choose a specific answer category based on their attributes, such as beliefs or socio-demographic characteristics.

In contrast to classical \emph{explanatory} regression models, such as ordinary least squares, where researchers are interested in controlling confounding factors, the objective here is to optimize \emph{predictive} performance. This means that the objective of predictive models is not to absorb the variation in potentially omitted variables that could bias the estimates for the explanatory variable(s) of interest as is the case in regression analysis. For instance, when trying to test the causal effect of urban-rural identity on support for climate policy, we may want to control for a third variable, the potentially confounding factor of income (it is not only urban-rural identity that drives support but also income). Urban voters typically have better opportunities in the labour market and earn higher incomes than rural voters. Wealth and higher income drive support for climate policy because environmental quality is a normal good, which means that its demand increases with income\cite{israel2004willingness}.

Although the difference between prediction and explanation may sometimes appear subtle to applied researchers, it reflects a critical division in empirical research: predictive classification models are data-driven and are evaluated based on their ability to predict an outcome correctly (here, public support for ambitious climate policy and political mobilization). Explanatory observational models that seek to test theories are typically evaluated theoretically regarding the extent to which they control for potentially confounding factors and provide estimates of unbiased effects that measure actual relationships.

Unlike ordinary least squares, random forest and XGBoost models use decision trees to make predictions. Based on the variables in the data, they split the data into increasingly small subsets until they only contain one class that is predicted to be the true class\cite{kern2019tree}. At each node of the tree, the data is split based on the predictor that contains the most information. This process optimizes the splits of the classes in the dependent variable -- in this analysis, meaning who supports or mobilizes around climate policy and who does not. 

\subsection{Model Performance and Selection}

\begin{figure}[!htb]
\centering
\includegraphics[width=.95\textwidth]{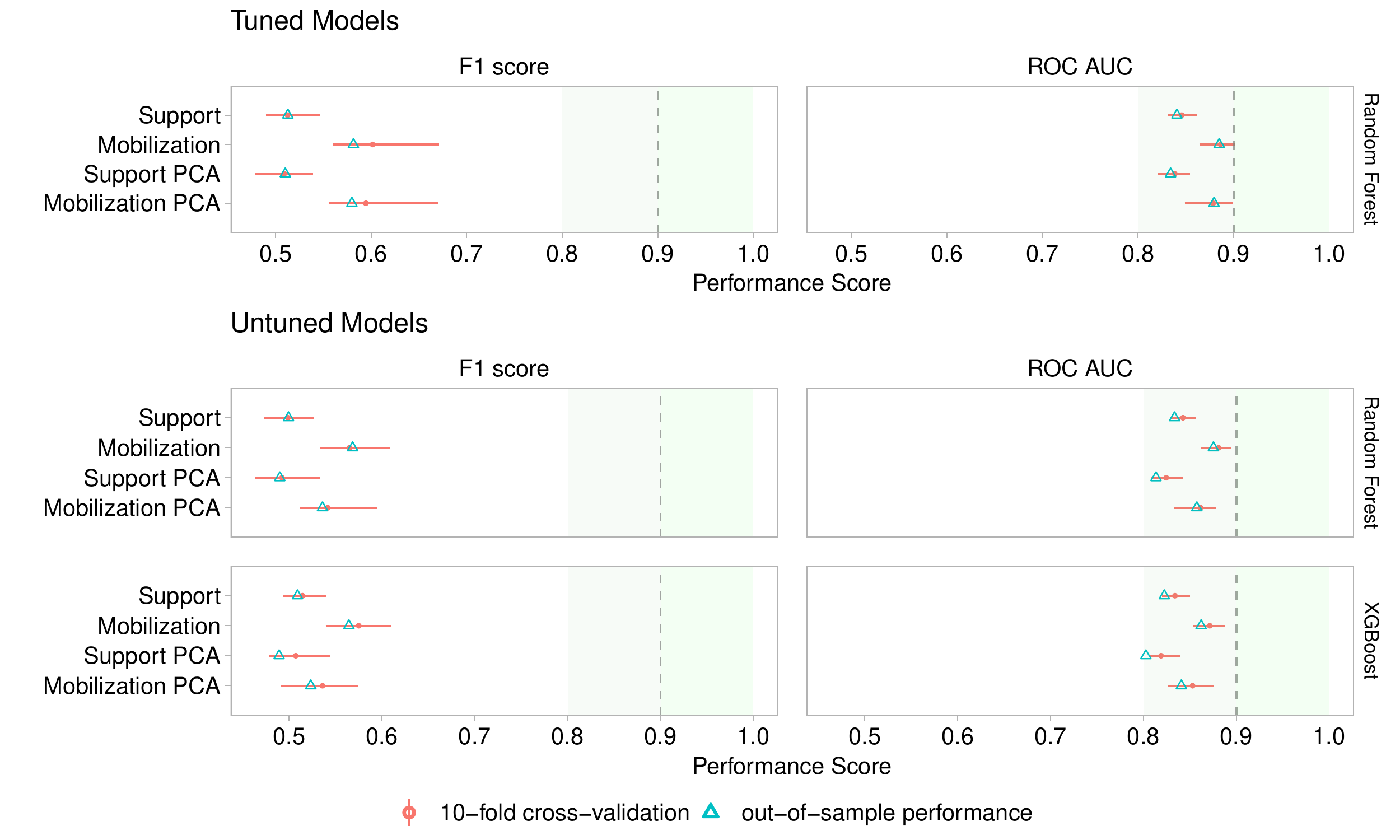}
\caption[Model performance]{Model performance using ROC area under the curve for public support and political mobilization using experimental and non-experimental data in Forest and XGBoost machine-learning models with and without PCA dimensionality reduction. Dots with error bars represent out-of-fold performance on the training data, and triangles represent out-of-sample performance.}
\label{fig:performance}
\end{figure}

To select the best-performing model, I evaluated the predictive models in Fig \ref{fig:performance}. First, to provide a holistic overview of the predictive importance of nine key variable groups established in the social science literature on public support for and political mobilization around climate policies (see Table \ref{tab:lit_overview}), I combined several survey questions within these overarching literature strands into one dimension. This can be done using a statistical method called principal component analysis (PCA; see, e.g., \cite{ringner2008principal}). Such dimensionality reduction simplifies the analysis and the understandability of the results. How well these variables can be reduced into one dimension is similar across variable groups (see Fig C in S1 Text): Each group absorbs at least 85\% of the total variance, and the share of the total variance is comparable across the different factors. This means that the difference in their predictive importance is not driven by how well they are reduced to one dimension. I also present results about each predictor to provide insight into which specific predictors of each principal component best predict the outcomes. Second, I compared two machine learning algorithms, the forest and the XGBoost model. This amounts to two outcomes (support and mobilization) $\times$ two operationalizations (PCA and individual variable effects) $\times$ two models (Forest and XGBoost), which equals eight models. In sum, these different models, specifications, and dependent variables provide robust insight into the performance of the models as well as the importance of the predictors. 

Machine-learning models are evaluated against data that was not used when the model learnt the relationships in the training data. This ensures that the model does not overfit (i.e., capture tendencies in the data that may not be important in other data), meaning that it is more generally applicable rather than data-specific. Data omitted whilst training, called test-data, is then used to evaluate model performance\cite{shmueli2010explain}. Model performance assessment on data not used during training is also called out-of-sample performance \cite{stone1974cross, geisser1975predictive, shmueli2010explain}. The respective scores may then be used to select the best-performing model for interpretation. This is a more restrictive approach to evaluating performance than using training data \cite{cranmer2017can}. Conventionally, 80\% of the data is used for training, and 20\% of the data is used to assess out-of-sample performance with tenfold cross-validation during training. Ten-fold cross-validation means that the training data is split randomly into ten subsets, each of which is omitted during training. Using cross-validation and out-of-sample prediction helps to prevent over-fitting of training data\cite{yadav2016analysis}. 

Fig \ref{fig:performance} shows the model performance for the random forest and the XGBoost model and two common model performance metrics, the F1 and receiver operating characteristic (ROC) area under the curve (AUC) multiclass scores. Higher scores indicate better performance. First, the F1 score shows the harmonic mean between precision and recall. Precision measures the proportion of correct positive classifications. Recall measures the extent to which the classifications miss predictions in the positive classes (that are instead predicted as negative; see Text B in S1 Text). 

\vspace{.3cm}

Second, the ROC AUC is a metric used to evaluate model performance. A perfectly optimized prediction model would classify all outcomes correctly. Overall performance concerning the rate of correct predictions can be measured by calculating the AUC, irrespective of the marginal uncertainty in ROC curves. The AUC statistic varies between 0.5, which indicates low predictive performance, and 1, which suggests that the model perfectly predicts the outcome.

Overall, both models achieve a similarly high F1 score. As the forest model is simpler and there is useful literature on the strengths and weaknesses of different variable importance scores introduced below, I tuned the hyperparameters of the random forest model. I used a space-filling grid search design for ten different hyperparameter combinations for the number of randomly sampled predictors at each tree split, the number of trees, and the data point minimum that was necessary to split the nodes into additional leaves (for the ROC AUC curves of the tuned models, see Fig D in S1 Text). The tuned model performed slightly better than the untuned model. The analysis and the interactive online tool which deploys the predictions rely on the tuned model.

\subsection{Variable Importance Measures}

Variable importance scores are a popular means of evaluating the predictive performance of machine learning models. The Gini variable importance measure was initially developed for the random forest approach\cite{breiman1984friedman}. Although these standard measures are widely used to assess predictive importance, they are biased. The importance of categorical or correlated predictors is often overestimated when using conventional Gini variable importance scores\cite{strobl2007bias}. 

Two variable importance measures are more robust than simple Gini measures. First, research has shown that permutation-based variable importance measures provide more reliable estimates of predictive importance than non-permutation-based approaches\cite{ziegler2014mining, nembrini2018revival}, such as the Gini variable importance measure. This is computed by subtracting the predictive performance score when all predictors are included with the score from the model when the predictor of interest is removed\cite{fisher2019all}. A more significant decline in the model's predictive performance indicates the greater predictive importance of that factor. Re-sampling the observed data multiple times provides estimates of the uncertainty associated with model performance, here displayed using violin plots in the results section in Fig \ref{fig:feature_imp_pca_forest_experimental} (see Text D in S1 Text for the algorithm). Second, bias correction algorithms for the Gini variable importance measure\cite{sandri2008bias} were implemented\cite{wright2017ranger}. Simulation studies validated that this method produces unbiased results\cite{loh2021variable}. 

I employed both these measures -- the bias-corrected approach\cite{sandri2008bias} and the permutation-based approach\cite{biecek2021explanatory} -- to generate robust and unbiased insight into the predictive importance of the variable groups. To further reduce sensitivity to different measurement scales, I normalized the data so that the predictors had a mean of zero and a standard deviation of one. To ensure that missing values did not distort the representative sample, I imputed them using the median value of the variable distribution. In addition, I employed robustness checks using an elastic net machine learning model\cite{zou2005regularization}, and OLS variable importance scores to ensure that the results were not driven by correlation in the predictors and further linear regression variable importance scores\cite{gromping2007relative, gromping2009variable}.

\section{Results}

\begin{figure}[!htb]
\centering
\includegraphics[width=.95\linewidth]{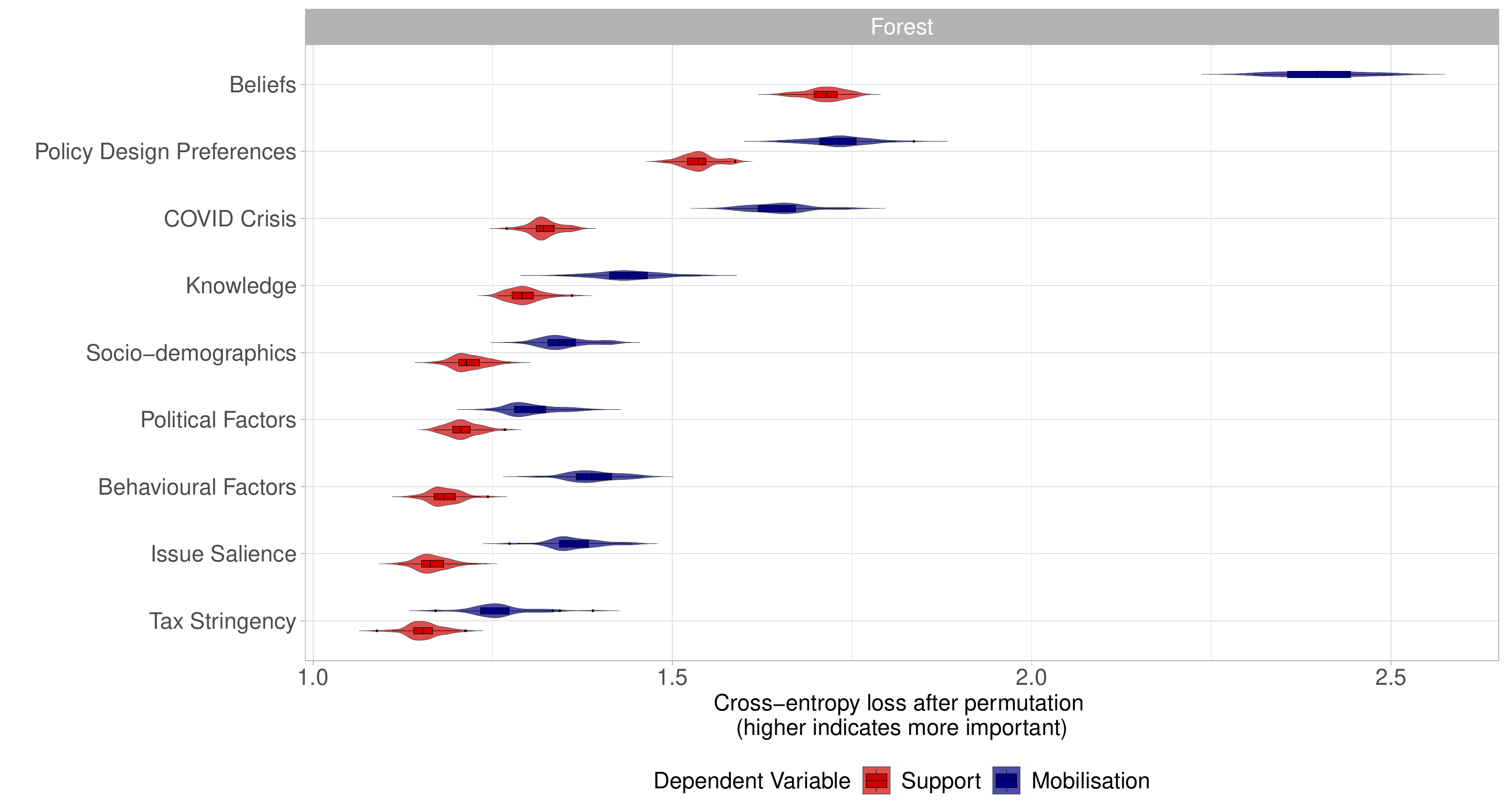}
\caption[Permutation-based variable importance scores using the tuned random forest model with PCA]{Permutation-based variable importance scores using the tuned random forest model with PCA. The box and violin plots display the distribution of cross-entropy loss for a given variable for 50 permuted samples from the observed data to account for uncertainty associated with predictions.}
\label{fig:feature_imp_pca_forest_experimental}
\end{figure}

To answer the research question, `\emph{What are the most important predictors of public support and political mobilization for ambitious climate policies?}', I proceed in two steps. First, I present results concerning the predictive power of the overarching variable groups from the literature introduced in the literature review (see Table \ref{tab:lit_overview}) using PCA in Fig \ref{fig:feature_imp_pca_forest_experimental} (for details, see methods). I use this technique to reduce several variables that are, according to the literature, associated with the same concept to obtain a general overview of their relative predictive power. This allows for connecting the findings on these concepts to established explanations in the literature and evaluating their relative predictive power holistically on the meta-level. Second, I provide nuanced results about the predictive power of the sub-concepts on the micro-level in Fig \ref{fig:feature_imp_pca_forest_experimental}. This more detailed approach shows what aspects of the larger variable groups drive variation in support and mobilization. 

\begin{figure}[!htb]
\centering
\includegraphics[width=.95\linewidth]{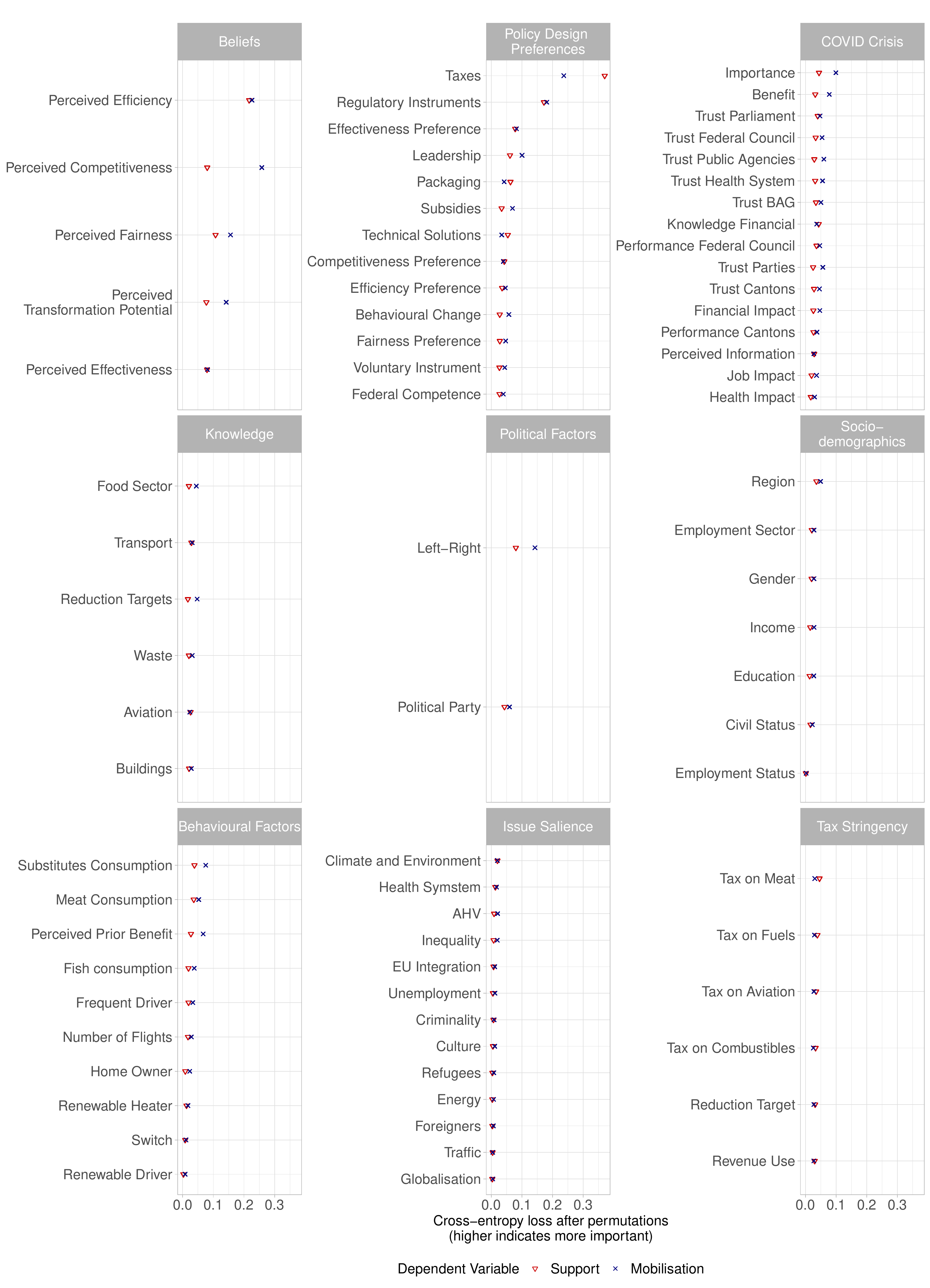}
\caption[Permutation-based variable importance scores using the tuned random forest model with all individual predictors]{Permutation-based variable importance scores using the tuned random forest model with all individual predictors. Dots represent the mean cross-entropy loss for a given variable for 50 permuted samples for the observed data to account for the uncertainty associated with the importance of predictions.}
\label{fig:feature_importance_experimental_forest}
\end{figure}

Overall, as shown in Fig \ref{fig:feature_imp_pca_forest_experimental}, the evidence indicates that beliefs (such as the perception of effects on the competitiveness of policies or their effectiveness at reducing CO$_2$ emissions) and policy design preferences concerning instrument choice (such as subsidy versus tax preferences) are the most important predictors of support and mobilization. This result is robust to other models and specifications, including elastic net and several different variable importance performance metrics (for details, see Text E as well as Fig F, Fig G, and Fig H in S1 Text). Beyond existing findings that beliefs are a statistically significant explanation for support for ambitious climate policy on their own right\cite{ingold2011network, kammermann2018beliefs, huber2020public, stadelmann2021public}, the results here show that they do have the highest predictive power \emph{relative} to all other factors in the model. More specifically, Fig \ref{fig:feature_importance_experimental_forest} shows that beliefs related to economic aspects (namely, economic efficiency and effects on economic competitiveness) are the most important sub-concepts. Thus, beliefs and the subjective evaluation of economic aspects are important drivers of support and mobilization. The third most important sub-concept captures if voters' beliefs that the CO$_2$ Act constitutes a just and fair policy with socially just distributional effects. Although still scoring high compared to the predictors in other groups, beliefs about the environmental impact of the CO$_2$ Act are less important than beliefs related to economic aspects of the law. 

Factors related to the COVID pandemic, such as the perception of the importance of the pandemic, trust in governance arrangements, knowledge about COVID-19-related regulations, and the impact of the crisis on individuals' health and economic situation, comprise the third important predictor. In contrast to beliefs and preferences, the predictive power of COVID-19-related factors is higher for political mobilization than for public support. This finding is robust to different variable importance scores (see Fig E in S1 Text using the Gini impurity-corrected variable importance measure\cite{sandri2008bias, loh2021variable}). These results show that cross-sectoral dynamics play an important role in climate and environmental policies, especially in terms of mobilizing voters for political participation. Political processes are not independent but interact and can be decisive in popular votes or elections, especially for voter turnout. 

Knowledge about key elements of the CO$_2$ Act, such as the reduction target and its sectoral coverage, is associated with greater predictive power for public support than for political mobilization. Specific design attributes, especially those related to the calibration of policy instruments such as revenue recycling schemes\cite{mildenberger2022limited}, often remain unnoticed. More knowledge about climate and what is on the bill is important for predicting support and mobilization. 

In contrast, other established explanations are associated with relatively weak predictive power. For example, political factors (e.g., left-right orientation and political party affiliation), sociodemographic variables, behavioural factors (e.g., meat consumption), salience (e.g., the perceived importance of climate and environment in comparison to other issues), and tax stringency in the housing, transport and food sectors have similarly weak predictive power. However, these findings do not imply that these factors are statistically and substantially non-significant explanations but instead imply that beliefs, policy design preferences and COVID-related factors matter more \emph{relative} to socio-demographics, political factors, behavioural factors, individual-level issue salience, and tax stringency for predicting public support and political mobilization.

The findings on political variables only partly support the claims of Hornsey et al.\cite{hornsey2016meta} who find from a meta-analysis across 56 countries that party identification with conservative ideology strongly predicts scepticism about climate change. The evidence here suggests that ideology and party affiliation in consensual political systems, such as that of Switzerland, are less likely to divide voters than in two-party political systems. Similarly, socio-demographic variables, such as income, region of residence, civil status, employment sector and gender, imply that traditional explanations for voting behaviour have relatively little predictive power. 

Behavioural factors, which include sectoral consumption behaviour (like consuming meat in the food sector, car driving in transport, or using renewable energy heating systems in the housing sector), tend to predict slightly more variation in political mobilization than in support. Behavioural factors determine the opportunity costs of behavioural change and are therefore important from a theoretical perspective in terms of informing potential pathways for policy change.

Issue salience, the perceived importance of climate and environment relative to other issues, is associated with relatively weak predictive power compared to beliefs and preferences. This finding is striking given its pervasive presence in the literature on voting behaviour in political science more generally\cite{repass1971issue, page1983effects, krosnick1990government, fournier2003issue, belanger2008issue, miller2017origins, bromley2020importance}. 

This finding for issue salience also has implications for normative theories of democracy\cite{dahl1956preface, page1983effects} and theories of policy change\cite{kingdon1984agendas, zohlnhofer2016bringing}. Normative theories of democracy hold that changes in the salience of issues should lead to policy changes\cite{dahl1956preface, page1983effects, stimson1995dynamic}. The results described here, however, suggest that this link is less clear than research suggests. Suppose issue salience affects the support of voters only to a minor extent. In this case, it is unlikely to create the degree of responsiveness that should be present according to normative theories of democracy. For issue salience to open windows of opportunity for policy entrepreneurs to push for policy change\cite{kingdon1984agendas, zohlnhofer2016bringing}, individual voter-level issue salience alone may not be enough. Rather, salience at the societal level may be more important for creating momentum for new policy proposals, especially in interaction with other favourable circumstances, such as individuals' beliefs and voter preferences or agenda-setting by elites. From the evidence presented here, it is however not entirely clear to what extent the popular vote influenced the salience of climate change and the environment and therefore the results. Further research should seek to confirm this finding in empirical settings that are not embedded in a popular vote. 

Last, tax stringency attributes also exhibit relatively little predictive power compared to the other factors. Although the evidence in the literature suggests that support can be increased by calibrating the design of carbon taxes\cite{drews2016explains, klenert2018making, mildenberger2022limited, bergquist2020combining}), choosing policy instruments other than taxes has a greater effect on support and mobilization than calibrating the design of the tax for instance through the inclusion of lump sum reimbursement.

\section{Conclusion}

Public support and political mobilization are two crucial components that shape the adoption and implementation of ambitious climate policies\cite{drews2016explains}. Ultimately, they will influence the feasibility of achieving the Paris Agreement targets for climate mitigation. To understand which solutions are politically feasible, research has so far primarily focused on explaining rather than predicting\cite{shmueli2010explain, cranmer2017can, beiser2018assessing} public support \cite{stokes2016electoral, drews2016explains, huber2020public} and mobilization \cite{mildenberger2019households, stokes2016electoral}. Little research, however, has focused on predicting the two. 

This study makes three contributions. First, predictive modelling can provide complementary insight into the relative importance of established explanations\cite{cranmer2017can}. This advances the theoretical debate concerning potentially competing theories and can contribute to theory building in subsequent research\cite{shmueli2010explain}. The results described in this study show that beliefs and policy design preferences are the most important predictors while other established explanations, such as socio-demographics, issue salience (the perceived relative importance of issues)\cite{repass1971issue, page1983effects, krosnick1990government, fournier2003issue, belanger2008issue, miller2017origins} or political variables (such as the party affiliation) have \emph{relatively} weak predictive power. However, these results do not indicate that the effect of these variables are substantially or statistically insignificant; they show that explanations concerning beliefs\cite{ingold2011network, kammermann2018beliefs, huber2020public, stadelmann2021public} and preferences\cite{kachi2015climate} matter more \emph{relative} to issue salience or specific policy instrument design calibration (such as the size of a tax). The finding that beliefs are a key predictor extends on prior research\cite{ kronsik2006origins, ingold2011network, huber2020public, gampfer2014individuals} by showing the predictive power of beliefs relative to other factors and not only on their own right. The strong predictive power of policy design preferences associated with instrument choices shows that the latter are important to consider in relation to the composition of coalitions\cite{metz2021policy} and thus the feasibility of achieving climate targets (see, e.g., \cite{drews2016explains}).

Second, predictive modelling can generate practically relevant knowledge. Accurate predictions can inform stakeholder strategies. The main model presented here generates reasonable predictive performance scores and may provide credible, policy-relevant insight. An interactive online tool is used to deploy the model. This allows the user to obtain real-time predictions whilst changing voter attributes. Thus, this model can be used to anticipate the support for carbon pricing policies of Swiss voters based on the predictors used in this model. 

Third, this study analyses two environmentally significant forms of behaviour in the public sphere: public support and political mobilization. The analysis is based on the centrepiece of Switzerland's climate policies, the Swiss CO$_2$ Act, and uses data from a repeated randomized forced-choice conjoint survey experiment. The survey experiment reduces social desirability bias. While the results show that the predictive power of the different variable groups broadly follows the same tendencies across the two outcomes, one key difference is worth noting: there is some evidence that cross-sectoral effects due to a crisis such as the COVID-19 pandemic matter more for political mobilization than for public support. These results show that climate policy processes are cross-sectoral: dependencies across domains can determine success or failure at the ballot box.

Despite the strong external validity of conjoint survey experiments in relation to replicating real-world behaviour\cite{hainmueller2014causal, hainmueller2015validating}, the specific focus on Swiss climate policy somewhat limits the external validity of the findings to other countries, especially to non-democratic or less developed countries. Therefore, research should analyze the power of predictors in different countries and, potentially, without the context of direct-democratic popular voting. This may also improve the external validity of the findings for countries without direct-democratic procedures. Such research could also be comparative in nature to improve understanding of how ongoing political processes such as the vote on the Swiss CO$_2$ Act affect issue salience and the finding presented in this study that this factor has comparatively little predictive power in respect of public support and mobilization. Public opinion research is critical to informing policymakers since public support is a crucial factor in adopting ambitious climate policies to limit global warming to below 2\textdegree. Therefore, research should produce to even more comprehensive online tools to inform policymakers about feasible policy proposals based on comparative multi-country data in varying policy process contexts, potentially over time. 

\section*{Financial Disclosure}

The author would like to acknowledge funding through a Swiss National Science Foundation grant in the Doc.CH funding scheme (SNF grant number 207269). This grant covered the author's salary as a Doctoral Candidate at the University of Bern while conducting this study. Data collection was supported by the SNF grant 188950. The funders had no role in study design, data collection and analysis, decision to publish, or preparation of the manuscript.

\section*{Competing Interests}

The author has declared that no competing interests exist.

\section*{Supporting information}

This analysis is supported by an interactive online application which can be accessed under \url{https://prediction.adibilis.ch}. Replication files can be accessed under \url{https://gitlab.adibilis.ch/rprojects/ShinyAppPredictions}.

\newpage

\nolinenumbers

%
%
%
%
%
%
%
\bibliography{sample}

\end{document}